\input amstex
\documentstyle{amsppt}
\loadbold
\magnification=\magstep1
{\catcode`\@=11\gdef\logo@{}}
\pagewidth{16 true cm}
\pageheight{25 true cm}
\voffset 0.5 cm
\document
\parindent = 0.5cm
\parskip = 1.5mm
\define\aaa{1.0cm}
\define\bbb{0.8cm}

\define\pa{\partial}

\define\pd#1#2{\frac{\pa#1}{\pa#2}}
\define\po#1{\frac{\pa}{\pa#1}}

\pageno=1
\NoRunningHeads
\TagsOnRight

\vskip 5cm
\centerline{\bf A LINEAR-CONFINED PARTICLE AND THE DIRAC EQUATION}
\vskip 1cm
\centerline{J. Weiss
\footnote{e-mail: weiss\@cvt.stuba.sk}}
\vskip 2cm
\centerline{\sl Department of Physics, University of Saints Cyril and
Methodius} 
\centerline{\sl Nam. J. Herdu 2, \,  917 01 Trnava, Slovakia}

\vskip 3cm
\indent
      The model of a classical particle with the weak linear AAD potential is
subjected to path integral quantization. The light cone constraints and 
peculiar properties of its internal variables permit to use in
calculations commutative dynamics and apply path integrals for a
matrix form of the transition amplitude. Quantization leads to description of 
a Dirac particle.
\bigskip
\bigskip
\bigskip
\vskip 4cm
\leftline{PACS: 03.65Pm, 11.25-w, 12.15-y, 11.15Ex}
\newpage
\vskip \aaa
\centerline{\bf 1. Introduction}
\vskip \bbb
   Motivated by the Wheeler-Feynman action-at-a-distance (AAD)
electrodynamics, a model of particle with the weak linear potential
was developed [1],[3], which allows to make reasonable physical
implications linking elements of its commutative dynamics with the light
cone constraints. Due to these constraints, there occurs the possibility
to performe relatively easily quantization of the model in the intention of
Feynman. The quantization requires to fulfil the important condition:
the internal variables of the model, inputted by AAD dynamics, have to
be regarded as a canonical pair. In this case they become a specimen
element of path integrals, which admits to introduce into the Feynman
formalism chains of Dirac  $\delta$ functions and thereby to facilitate
calculations. Beside the use of the internal variables as the tools
of pure Hamiltonian dynamics, it will be emphasized in this paper even
the role of the Routh function in the dynamical considerations. Its
introduction consists in the effort to give a more general meaning
to these variables in the process of quantization. They could express,
in our opinion, the connection between their light-cone constraint
content and spinning of particle; at least to that extent as
it was anticipated by Feynman [2], when he tried to establish a compromise
solution of spin 1/2 particle dynamics.

In the paper [3] we have formulated major features of relativistic
classical dynamics of the point particle with the internal degrees of
freedom, defined according to the AAD theory of linear interaction taken for
weak coupling. Now we want to show that this model leads to quantum dynamics
of a particle with the fixed mass and the spin 1/2. Dynamics of this
particle must be supplemented by subsidiary conditions, which are,
however, consistent with the equations of motions. Moreover, they are
sufficient to construct the Feynman integrals along the paths.
The appropriate quantum equation, obtained in this formalism,
is the equation for the two-component spinor, proposed by Feynman
and Gell-Mann [4], which is equivalent to the Dirac equation. The
nonrelativistic Feynmanian quantization of the model is given in [7].

In Sec. 2 a brief summary of main properties of classical model is given. 
Sec. 3 explains that if the constraints are considered to involve even virtual
motions and the presence of weak electromagnetic action, they are more
flexible, when the Routh function is introduced. In Sec. 4 it is argued that the
Feynman formalism demands to adopt four transition amplitudes and to
add some subsidiary conditions in close connection with the
constraints and the equations of motion. The form, in which the internal
variables must be represented to correspond to the complete structure of
transition amplitude, appears to have the spinor behaviour.
\vskip \aaa
\centerline{\bf 2. Glossary of the classical model }
\vskip \bbb
   Before discussing the quantum version of the model
let us summarize major properties of its classical picture.
The examinated particle was described [1] as a point object with internal degrees
of freedom, dynamics of which is related to the transitive realizations
of Poincar\'e' s group with the generators
$$M^{\mu\nu} = x^{\mu}p^{\nu} - x^{\nu}p^{\mu} + \xi^{\mu}\eta^{\nu} -
\xi^{\nu}\eta^{\mu}, \eqno(2.1)$$ where $\xi$ and $\eta$ were considered as 
the canonically conjugate internal variables (properly $\xi$ and $b\eta$
evaluated with $b = 1$) obeying the constraints 
$$\xi^{2} = 0\, \quad \quad \eta^{2} = 0\,. \eqno(2.2)$$
The physically most interesting are realizations that correspond to the
case $p^{2} \geq 0$. In this case the irreducible unitary representation
of small Lorentz's group is finitely dimensional. Then one can avoid 
calculations with $\xi$ and $\eta$ as the continual variables and pass
straightforwardly to matrices, which is a standard approach resulting into
conventional matrix description of spin [2].

Hence, we utilize the variables $\xi$ and $\eta$ for all region $p^{2} \geq $.
In order that the realization of the Poincar\'e group remains transitive,
we subject these variables beside (2.2) the following conditions
$$(\xi. p) = \kappa \,\quad  (\eta. p) = \kappa' \eqno (2.3a)$$
$$(\xi. \eta)p^{2} = 2(p.\xi)(p.\eta) = 2\kappa\kappa',  \eqno(2.3b)$$
$\kappa$ and $\kappa'$ being some constants. Such a particle is possible
then to be conceived as an elementary object, deduced from the form of linear
AAD interaction, having all necessary atributes of conventional particles, 
\it i.e. \rm $(m,\vec p,\vec x)$. Moreover, its state can be characterized
by one unity vector $\vec n = \vec \xi
/\xi^{0}$. The two degrees of freedom, peculiar to this vector, give the
direct physical content for the quantum picture of the particle.

The particle fulfils the variation principle
$$\delta \Bigl \{ - \int_{\tau_{1}}^{\tau_{2}} d \tau [(\eta.\dot \xi) + 
R] + (\xi.\eta)_{\tau = \tau_{2}}\Bigr \} = 0 \,  ,\eqno(2.4)$$ 
where $R$ is the Routh function and the constraints (2.2) and (2.3) are
valid. The equations of motion, derived from (2.4), are
$${d\over d\tau}{\partial R\over \partial u^{\mu}} - {\partial R\over
\partial x^{\mu}} = 0 $$
$${d\vec \xi\over d \tau} = {\partial R\over \partial \vec \eta} + 
{\vec \eta\over (\xi.\eta)}\Bigl ( \vec \xi. {\partial R\over \partial \vec 
\eta}\Bigr )\,; \quad {d\vec \eta\over d \tau} = -\Bigl [{\partial R\over 
\partial \vec \xi} + {\vec \xi\over (\xi.\eta)}\Bigl (\vec \eta.{\partial 
R\over \partial \vec \xi}\Bigr ) \Bigr ]$$
$${d\xi^{0}\over d \tau} = {\eta^{0}\over (\xi.\eta)}\Bigl (\vec \xi.
{\partial R\over \vec \eta}\Bigr ) \,;\quad {d\eta^{0}\over d \tau} = 
-{\xi^{0}\over (\xi.\eta)}\Bigl (\vec \eta.{\partial R \over \partial \vec \xi}
\Bigr )\,. \eqno(2.5)$$ 

One expects naturally that dynamics of the free particle will be
sufficiently simple. We can also believe that on the classical level the
fourvectors $\xi$ and $\eta$ will refer to a "residuum" of particle spin
according to intuition of Feynman. This spin relic, of course, must be
a constant in time without the presence of electromagnetic field [2].
Therefore it is natural to assume for the free particle to be
$${d\xi^{\mu}\over d\tau} = 0\, ; \quad \quad {d\eta^{\mu}\over d\tau} = 0\,.
\eqno(2.6)$$  
The Routh function, independent in this case on $x$, $\xi$ and $\eta$, and
expressed in terms of the fourvelocity $u$, is given by
$$R = {1\over 2}\Bigl (\tilde \mu u^{2} + {m^{2}\over \tilde \mu}\Bigr )
\, \eqno(2.7) $$
where $\tilde \mu$ is an auxiliary variable introduced for the action $S$ to be
invariant under the reparametrization. It is obvious that for the momentum
we have $p^{\mu} = {\partial R\over \partial u^{\mu}}$ ; 
${dp^{\mu}\over d \tau} = 0 $ and $p^{\mu} = \tilde \mu u^{\mu}$. 
It is also evident that the constraints (2.3) are consistent with the
equations of motion of this free particle.

Note that the conditions (2.3), considered either for $p$ or $u$, admit
in fact only two degrees of freedom, associated \it e.g. \rm with the 
orientation of $\vec \xi$. Furthermore, for $\xi^{\mu}$ it holds $\xi^{2} =
0$ and $(\xi.p) = \kappa$, which admits any direction for $\vec \xi/
\xi^{0}$.

Let us give the form of the action functional that leads directly to the
Hamilton canonical equations. It is
$$S = - \int_{\tau_{1}}^{\tau_{2}}d\tau \Bigl [(p.u) + (\eta. \dot \xi) +
H \Bigr ] + (\xi.\eta)_{\tau=\tau_{2}} \, , \eqno(2.8)$$
where $$H = {1\over 2\tilde \mu}(m^{2} - p^{2}) \, . \eqno(2.9)$$
The Hamilton-Jacobi function $S_{21}$ can be evaluated as $S$ taken for 
real motion. We have
$$S_{21} = - {1\over 2}\Bigl [{(x_{2} - x_{1})^{2}\over \rho_{21}} + m^{2}
\rho_{21}\Bigr ] \, , \eqno(2.10)$$
$\rho$ being defined as $d\rho = {d\tau\over \tilde \mu}$.
In (2.10)
the dependence of $S_{21}$ on $\xi$ and $\eta$ is not present
because $(\eta.\dot 
\xi) = 0$ and the term $(\xi.\eta)_{\tau=\tau_{2}}$ yields a constant
factor due to (2.3). The form of the action $S_{21}$ is now
$$S_{21} = - Extr \Bigl \{[p_{21}.(x_2 - x_1)] + H_{21}\rho_{21} \Bigr \} \, , 
\eqno(2.11)$$ where now $$H_{21} = {1\over 2}(m^{2} - p^{2}_{21})\,.$$

The behaviour of the model if the electromagnetic field is present is 
described by the Routh function 
$$R = {1\over 2}\Bigl (\tilde \mu u^{2} + {m^{2}\over \tilde \mu} \Bigr ) +
e\tilde A_{\mu}u^{\mu} + \lambda F_{\mu\nu}\xi^{\mu}\eta^{\nu} \, , \eqno
(2.12)$$ where $\lambda$ is a coupling constant and $\tilde A_{\mu}$ 
and $F_{\mu \nu}$ are the fourpotential and strenght of electromagnetic field,
respectively. $\lambda$ can be determined by the requirement for the 
constraints $\tilde \mu(\xi.u) = \kappa$ and $\tilde \mu(\eta.u) = \kappa'$ to
be in agreement with the equations of motion. Of course, they are
$${d\over d\tau}\Bigl (\tilde \mu u^{\mu}\Bigr )= eF^{\mu \nu}u^{\nu} ,\eqno
(2.13)$$ with $F^{\mu \nu}=\tilde A_{\nu,\mu}-\tilde A_{\mu,\nu}.$  
We assume, again as in the case of linear field, that action of
electromagnetic field is weak, omitting the term $\partial _{\mu}(F_{\rho \nu}
\xi^{\rho}\eta^{\nu}).$ 

The equations of motion for $\xi$ and $\eta$ are the consequences of the
variation principle. We must here evaluate the corresponding derivates
of $R$ over $\vec \xi$ and $\vec \eta$. The result is simple:
$${d\xi^{\mu}\over d\tau}=-\lambda F^{\mu \nu}\xi_{\nu}\, ; \quad
{d\eta^{\mu}\over d\tau}=\lambda F^{\mu \nu}\eta_{\nu},$$
with $\lambda=e/\tilde \mu$ deduced from the condition $\tilde
\mu(\xi.u)=const$.

It can be shown that the conditions (2.3) are consistent with the condition
$${\xi^{\mu}\over \tilde \mu(u.\xi)} + {\eta^{\mu}\over \tilde \mu(u.\eta)}
- 2{u^{\mu}\over \tilde \mu u^{2}} = 0 \, , \eqno(2.14)$$
and thus with the equations of motion. Eq.(2.14) may be viewed physically 
as the fact that the external field is not able to cause spin excitations
of the particle. 

The Hamilton formalism requires now for $p$ and $H$ the following forms
$$p^{\mu} = \tilde \mu u^{\mu} + e\tilde A^{\mu} \, \eqno(2.15)$$
$$H = {1\over 2\tilde \mu}\Bigl [m^{2} - (p^{\mu} - e\tilde A^{\mu})^{2}\Bigr ]
+ {e\over \tilde \mu} F_{\mu \nu}\xi^{\mu}\eta^{\nu} \, , \eqno(2.16)$$
respectively, or using the variable $\rho$, equivalently
$$H = {1\over 2}\Bigl [ m^{2} - (p^{\mu} - e\tilde A^{\mu})^{2}\Bigr ] +
eF_{\mu \nu} \xi^{\mu}\eta^{\nu} \, . \eqno(2.17)$$
It can be easily verified that the Hamilton-Jacobi function depends only
on $x$ and $x'$ , but not on $\xi$ and $\eta$. 
As a result the action integral is identical with that defined
for the scalar particle.  

Dynamics for the variables $\xi$ and $\eta$ is thus determined, if 
$F_{\mu \nu} \not = 0$, by the same subsidiary conditions as that for the
free particle. This means, for the same times $\rho$ the condition
(2.14) will be identical with that for the free particle. For the different
$\rho$ it will differ.
\vskip
\aaa
\centerline{\bf 3. Quantum dynamics of the model}
\vskip
\bbb
   In the Feynman formalism dynamics of a particle is determined by the
transition amplitude from the state at the "time" $\rho_{1}$, to the state
at the "time" $\rho_{2}$. Our model is accomodated to have four amplitudes
$$A_{21} = A_{21}(x_{2},x_{1},\xi_{2},\xi_{1},\rho_{21}) \, ; \quad B_{21}
= B_{21}(x_{2},x_{1},\eta_{2},\eta_{1},\rho_{21})$$
$$C_{21} = C_{21}(x_{2},x_{1},\eta_{2},\xi_{1},\rho_{21}) \, ; \quad
D_{21} = D_{21}(x_{2},x_{1},\xi_{2},\eta_{1},\rho_{21})\, . \eqno(3.1)$$
The structure of these amplitudes demands their dependence on $x_{2}$ and
$x_{1}$ to be defined by the standard exponential of the type
$$\rho_{21}^{-2}e^{-{i\over 2\hbar}[\rho_{21}^{-1}(x_2-x_1)^{2} + 
m^{2}\rho_{21}]}
\, .\eqno(3.2)$$ On the other hand, the dependence on $\xi$
and $\eta$ have to be characterized by Dirac's delta functions in a way by
which the subsidiary conditions for $\xi$ and $\eta$ are suitably
expressed (relations (2.3) or (2.14)). 

The integrals over trajectories are usually taken in the form involving
the integration over momenta. Therefore we shall use the following integrals
$$\int{ e^{{i\over \hbar}\{-[p_{21}.(x_2-x_1)] + {1\over
2}(p^{2}_{21} - m^{2})\rho_{21}\}}{d^{4}p_{21}\over (2\pi \hbar)^{4}}} = $$
$$= {1\over i(2\pi \hbar)^{2}\rho^{2}_{21}}{e^{-{i\over 2\hbar} [\rho^{-1}_{21}
(x_2-x_1)^{2} + m^{2}\rho_{21}]}} \,  .            \eqno(3.3)$$
The continual integration over $\xi$ and $\eta$ will be represented by the
differentials
$${d^3\eta_2\over \eta_2^0} \,{d^3\xi_2\over \xi_2^0} \, {d^3\eta_3\over
\eta^0_3} \, {d^3\xi_3\over \xi^0_3}   .....$$
which are the Lorentz invariant quantities. One of kinds of auxiliary
conditions will be expressed by the combinations of $\delta$ functions:
$$\delta [(\xi_{1}.p_{21}) - \kappa]\delta [(\eta_{2}.p_{21}) - \kappa']
\delta [(\xi_{2}.p_{32}) - \kappa]\delta[(\xi_{3}.p_{32}) - \kappa'] 
 ....
$$

If one introduces into (2.3b)
$$\tilde \xi^{\mu} = \xi^{\mu} - 2p^{-2}p^{\mu}(p.\xi) \, ;\quad \tilde
\eta^{\mu} = \eta^{\mu} - 2p^{-2}p^{\mu}(p.\eta) \, , \eqno(3.4)$$
then the subsidiary conditions may be written, respectively
$$(\tilde \xi.\eta) = 0 \, ;\quad (\tilde \eta.\xi) = 0 \,. \eqno(3.5)$$
So we find straightforwardly the analogon of conditions (2.2)
$$ \tilde \xi^{2} = 0 \, ; \quad \quad \tilde \eta^{2} = 0 \, .
\eqno(3.6)$$
It is apparent that the relations (3.5) lead to the new conditions
$$ \delta[(\tilde \xi.\eta)] = \delta[(\xi.\tilde \eta)] \, , \eqno(3.7)$$
expressed in terms of $\delta$ functions. As a result, 
in the functional 
integral there appears a $\delta$ function chain of the form
$$\delta [(\tilde {\xi_1}.\eta_2)] \delta[(\eta_2.\tilde {\xi_2})]
\delta [(\tilde {\xi_2}.\eta_3)] \delta [(\eta_3.\tilde {\xi_3})]
\delta [(\tilde {\xi_3}.\eta_4)] \delta [(\eta_4.\tilde {\xi_4})]
 .... \eqno(3.8)$$
Let us calculate, for instance, the integral
$$\int {\delta [(\tilde {\xi_{1}}.\eta_{2})]\delta [(\eta_{2}.\tilde 
{\xi_{2}})] \kappa' \delta [(\eta_{2}.p_{21}) - \kappa']{d^{3}\eta_{2}\over
\eta^{0}_{2}}} \, ,\eqno(3.9)$$
to find the way for verification of composition law. Due to relations
$$\kappa'\delta [(\eta_{2}.p_{21}) - \kappa'] = {\kappa'\over p^{0}_{21} - 
\vec n'_{2}.\vec p_{21}}\delta \Bigl (\eta^{0}_{2} - {\kappa'\over p^{0}_{21}
 - \vec n'_{2}.\vec p_{21}}\Bigr ) \, ,\eqno(3.10) $$
where $$\delta [(\eta_{2}.\tilde {\xi_{2}})] = (\eta_{2}^{0} \tilde 
{\xi^{0}_{2}})^{-1} \delta (1 - \vec n_{2}'.\tilde { \vec n_{2}}) \,; 
\quad  \delta [(\tilde {\xi_{1}}.\eta_{2})] = (\tilde {\eta_{1}^{0}}
\eta_{2}^{0})^{-1} \delta (1 - \tilde {\vec n_{1}}.\vec n'_{2})\, ,$$
and we put $\tilde {\vec n_{1}} = \tilde {\vec \xi_{1}}/\tilde
{\xi^{0}_{1}}$ \, ;\, $\tilde {\vec n_{2}} = \tilde {\vec \xi_{2}}/\tilde
{\xi^{0}_{2}}$ \, ;\, $ \tilde {\vec n^{2}_{1}} = 1$ \, ;\, $\tilde 
{\vec n^{2}_{2}} = 1$. Then the integral (3.9) yields
$$\int d\Omega_{\vec n'_{2}}(\tilde {\xi^{0}_{1}}\tilde {\xi^{0}_{2}})^{-1}
\delta (1 - \vec n'_{1}.\tilde {\vec n_{2}})\delta (1 - \tilde {\vec n_{1}}.
\vec n'_{2}) = 2\pi \delta [(\tilde {\xi_{1}}.\tilde {\xi_{2}})]\,. 
\eqno(3.11)$$ However, since $(\tilde {\xi_{1}}.\tilde {\xi_{2}}) = (\xi_{1}.
\xi_{2})$, one has finally
$$\int {1\over 2\pi}\delta [(\tilde {\xi_{1}}.\eta_{2})] {1\over 2\pi} 
\delta[(\eta_{2}.\tilde {\xi_{2}})] \kappa'\delta [(\eta_{2}.p_{21}) - \kappa']
{d^{3}\eta_{2}\over \eta^{0}_{2}} = {1\over 2\pi}\delta [(\xi_{1}.\xi_{2})]. 
\eqno(3.12)$$ So we see that the validity of the composition law requires to
accept in the role of coeficients staying before the $\delta$ functions the 
factors ${1\over 2\pi}$ \, , $\kappa$ and $\kappa'$, respectively. 
In a similar way we can compute integrals of other combinations for arguments
of $\delta$. 

Introduce the symbolical denotation

$$\Cal D x(\rho) = d^4x_2d^4x_3d^4x_4......d^4x_{N-1}\, ; \quad
\Cal D p(\rho) = {d^4p_{21}\over
(2\pi \hbar)^4}{d^4p_{32}\over (2\pi \hbar)^4}.....{d^4p_{N,N-1}\over 
2\pi \hbar)^4}$$
$$\Cal D \xi(\rho) = {d^3\xi_2\over \xi^0_2}{d^3\xi_3\over \xi^0_3}....
{d^3\xi_{N-1}\over \xi^0_{N-1}}\, ; \quad
 \Cal D \eta(\rho) = {d^3\eta_2\over \eta^0_2}{d^3\eta_3\over \eta^0_3}
......{d^3\eta_{N-1}\over \eta^0_{N-1}} \, .\eqno(3.13)$$
The amplitudes acquire then to have the form
$$A_{N1} = \int e^{{i\over \hbar}\sum_k S_{k+1,k}}\delta [\xi(\rho),\xi(\rho)]
\Cal D p(\rho)\Cal D x(\rho)  \Cal D \xi(\rho)  \Cal D \eta(\rho)
\eqno(3.14)$$ \rm
and analogically for $B_{N1}, C_{N1}$ and $D_{N1}$.

The integration over $x$ and $p$ in the composition law can be performed
in the standard way and reads
$$\int {d^4p_{21}\over (2\pi \hbar)^4}\int {d^4p_{32}\over (2\pi \hbar)^4}
\int \Bigl [e^{{i\over \hbar}\{-[p_{21}.(x_2-x_1)+p_{32}.(x_3-x_2)]+
{1\over 2}[(p_{21}^2-m^2)\rho_{21}+(p_{32}^2-m^2)\rho_{32}]\}}\Bigr ]
\times$$
$$ d^4x_2 = \int {d^4p_{31}\over (2\pi \hbar)^4}e^{{i\over \hbar}\{-[p_{31}.
(x_3-x_1)] + {1\over 2}(p_{31}^2-m^2)\rho_{31}\}} \, , \eqno(3.15)$$
where $\rho_{31} = \rho_{32}+\rho_{21}.$ We see that due to the
integration over $x$ the neighbouring momenta $p_{21}$ and $p_{32}$ are always
equal. Therefore it is irrelevant what momentum was used to define $\tilde
{\xi}$ and $\tilde {\eta}$. However the ambiguity of selecting $p$ plays no
role.
\vskip
\aaa
\centerline{\bf 4. Quantum model and spinors}
\vskip
\bbb
   The previous analysis has shown that we can make
a selection of four amplitudes and also that we have the
possibility to use a matrix form of the amplitude with the 2x2 dimensions.
The complete structure of the propagator must of course be of the type 
consisting of 4x4 Dirac's $\gamma$ matrices. Therefore we adjoint the undot
spinor $\zeta$ to each $\xi^\mu$, namely
$$(\xi^0 - \vec \sigma .\vec \xi)\zeta = 0 \, , \eqno(4.1)$$
where $\vec \sigma$ are the Pauli matrices. The solution of Eq.(4.1) is
$$\left(\matrix \zeta_{1}\\ \zeta_{2} \endmatrix\right)
 = c\left(\matrix e^{-i{\varphi\over 2}}\cos {\vartheta\over 2} \\
e^{i{\varphi\over 2}}\sin {\vartheta\over 2} \endmatrix\right)
 \eqno(4.2)$$
where $\vartheta$ and $\varphi$ are the spherical angles of the unity vector
$\vec n$, and $c$ a normalization factor. The spinor $\zeta$, normalized
so that 
$$\xi^\mu = (\zeta^+\zeta,\zeta^+\vec \sigma \zeta)\, ,  \eqno(4.3)$$
yields $\mid c\mid = \sqrt {\xi^0}$. The phase may be choosen arbitrarily
and so we chose $c=\mid c \mid$. Note that we can refer $\zeta$ to $\xi^\mu$
unambiguosly only if the condition (2.2) is obeyed.

Likewise we can introduce for each $\eta$ the dot spinor $\chi$, using the
equation
$$(\eta^0 + \vec \sigma. \vec \eta)\chi = 0 \,, \eqno(4.4)$$
the solution of which (now $\eta^2 =0$) is
$$\chi = \left(\matrix \chi_{\dot 1} \\
\chi_{\dot 2} \endmatrix\right) = c'\left(\matrix
e^{-i{\varphi\over 2}}\sin {\vartheta\over 2} \\
e^{i{\varphi\over 2}}\cos {\vartheta\over 2} \endmatrix\right)
 \eqno(4.5)$$
this time with the spherical angles of the vector $\vec n=\vec \eta/\eta^0$.
Normalizing $\chi$ according to
$$\eta^\mu = (\chi^+\chi, - \chi^+\vec \sigma \chi) \, , \eqno(4.6)$$
we obtain $\mid c' \mid = \sqrt {\eta^0}$.

The relative phase of $\chi$ under $\zeta$ is deduced from the requirement
to assure the transition $\zeta \rightarrow \xi$ using the space inversions
$\vec \xi \rightarrow -\vec \xi$ and $\vec \eta \rightarrow -\vec \eta$,
because in this case Eq.(4.1) passes to Eq.(4.4) and vice versa. Due to
the inversions $\theta \rightarrow \pi - \theta$ and $\phi \rightarrow 
\phi + \pi$, Eq.(4.2) acquires the fashion
$$\zeta = i\sqrt {\xi^0} \left(\matrix
e^{-i{\varphi\over 2}}\sin {\vartheta\over 2} \\
e^{i{\varphi\over 2}}\cos {\vartheta\over 2} \endmatrix\right)
. \eqno(4.7)$$
Next we adopt $\chi$ as follows
$$\chi = i\sqrt {\eta^0} \left(\matrix
-e^{-i{\varphi\over 2}}\sin {\vartheta\over 2} \\
e^{i{\varphi\over 2}}\cos {\vartheta\over 2} \endmatrix\right)
 \eqno(4.8)$$
and at the same time we secure the validity   
$${\vec \xi\over \xi^0} = - {\vec \eta\over \eta^0}\, ;\quad (\xi^0)^{-1/2}
\zeta = (\eta^0)^{-1/2}\chi \, . \eqno(4.9)$$ This simple equation suits
to the situation, when $\vec p = 0$, since in this case from (2.3) it follows
(4.9), as well $p_0\xi^0 = \kappa$\, ; \, $p_0\eta^0 = \kappa'$. It means
we have for $\vec p = 0$: $(\kappa)^{-1/2}\zeta = (\kappa')^{-1/2}\chi$.
For $\vec p \not = 0$ we again use (2.3) and derive
$$p^{-2}(p^0 - \vec \sigma. \vec p)\zeta = \chi'\zeta \, ;\quad
p^{-2}(p^0 + \vec \sigma. \vec p)\chi = \zeta' \chi \, , \eqno(4.10)$$
where 
$$\chi' = {p^2\over 2(p.\eta)}(\eta^0 - \vec \sigma.\vec \eta)\zeta\, ; \quad
\zeta' = {p^2\over 2(p.\xi)}(\xi^0 + \vec \sigma. \vec \xi)\chi \, . $$
The solutions of Eqs.(4.10) are unique and equal $ \xi' = c'\chi$ and
$\zeta' = c\zeta$ up to the two Lorentz invariant factors $c$ and $c'$.
It is easily verified for both the factors to be
$$c = \sqrt {p^2}{\sqrt {\kappa'}\over \sqrt \kappa} \, ; \quad 
c' = \sqrt {p^2}{\sqrt \kappa\over \sqrt {\kappa'}} \, . \eqno(4.11)$$
The ultimate form of Eqs.(4.10) is hence
$$(p^0 - \vec \sigma. \vec p)\zeta = {\sqrt {p^{2}\kappa}\over \sqrt {\kappa'}}
\chi \, ; \quad (p^0 + \vec \sigma. \vec p)\chi = {\sqrt {p^{2}\kappa'}\over
\sqrt \kappa}\zeta \, , \eqno(4.12)$$
being $$ (\eta^0 - \vec \sigma. \vec \eta)\zeta = 2{\sqrt {\kappa \kappa'}\over
\sqrt {p^2}}\chi \, ; \quad (\xi^0 + \vec \sigma. \vec \xi)\chi =
2{\sqrt {\kappa \kappa'}\over \sqrt {p^2}}\zeta \, .$$ 

Continual integrals over the variables $\xi$ and $\eta$ are possible to be
carried out just as in case of integration over $p$ in the previous section. 
Now, there remains in the transition amplitude only the $\delta$ function 
chain of types $\delta [(\xi_1.\xi_2)]$,
$\delta [(\eta_1.\eta_2)]$, or possibly $\delta [(\xi.\tilde {\eta})]$.

The integrals with the spinor products are coupled with the following form,
which gives for $p^2>0$
$$\int \delta[(\xi.p)-\kappa]{d^3\xi\over \xi^0} = \int {\kappa d
\Omega_{\vec n}\over (p^0-\vec n.\vec p)^2} = {4\pi \kappa\over p^2} \,. 
\eqno(4.13)$$
Using (4.13) these integrals read
$$\int \zeta \zeta^{+} \delta [(\xi.p) - \kappa]{d^3\xi\over \xi^0} =
2\pi{\kappa^{2}\over p^4}(p^0 + \vec \sigma.\vec p) \eqno(4.14a)$$
$$\int \chi \chi^{+} \delta [(\eta.p) - \kappa']{d^3\eta\over \eta^0} =
2\pi{\kappa^{'2}\over p^4}(p^0 - \vec \sigma.\vec p) \,. \eqno(4.14b)$$
We exploit now the results (4.14) and derive the appropriate amplitude
$$\int {d^4p\over (2\pi \hbar)^4} \int {d^3\xi_1\over \xi_{1}^{0}}
\int {d^3\xi_2\over \xi_{2}^{0}}e^{{i\over \hbar}S_{21}}\zeta_1 \zeta_{2}^+
\kappa \delta [(\xi_1.p)-\kappa].$$ $$.{1\over 2\pi}\delta [(\xi_1.\xi_2)]
\kappa \delta [(\xi_2.p)-\kappa] \, , \eqno(4.15)$$
where $\zeta_1$ refers to $\xi_1$ and $\zeta_2$ to $\xi_2$, respectively.
The amplitude (4.15) can be modified as follows
$$\int {d^4 p\over (2\pi \hbar)^4} \int {d^3\xi_1\over \xi_{1}^{0}}
e^{{i\over \hbar}S_{21}}\zeta_1\zeta_{1}^{+} \kappa \delta [(\xi_1.p)-\kappa]
\, , \eqno(4.16)$$ because due to $\xi_{1}^{0} = \kappa(p^0-\vec n_1.\vec
p)^{-1}$ we have 
$$\int {d^3\xi_{2}\over \xi_{2}^{0}}\zeta_{2}^{+} \delta [(\xi_1.\xi_2)]
\delta [(\xi_2.p)-\kappa] = {2\pi\over \kappa}\zeta_{1}^{+} \, .$$
Next, with help of (4.14a) the integral (4.16) yields
$$2\pi \int {d^4 p\over (2\pi \hbar)^4} e^{{i\over \hbar}S_{21}}\kappa^3
p^{-4}(p^0 + \vec \sigma.\vec p) \, . \eqno(4.17)$$
Note that a correct propagator should not have $p^{4}$
in the denominator. We must thus eliminate in (4.17) the quantity $p^{-4}$.
This may be made easily by the direct applying the operator $-\hbar
\partial_{1\mu}^{2} \partial_{2\mu}^{2}$ on the amplitude deduced. So
we have
$$2\pi \kappa^3 i\hbar\Bigl ({\partial \over \partial x_{2}^{0}} -
\vec \sigma. {\partial \over \partial \vec x_{2}}\Bigr ) \int {d^4p
\over (2\pi \hbar)^4} e^{{i\over \hbar}S_{21}} \, . \eqno(4.18)$$

By analogy with (4.15) one can create the amplitude associated with the 
variable $\eta$ 
$$\int {d^4 p\over (2\pi \hbar)^4} \int {d^3 \eta_1\over \eta_{1}^{0}}
\int {d^3 \eta_2\over \eta_{2}^{0}} e^{{i\over \hbar}S_{21}}\chi_{1} \chi_{2}^+
\kappa' \delta [(\eta_1.p)-\kappa'].$$ $$.{1\over 2\pi}\delta [(\eta_1.\eta_2)]
\kappa' \delta [(\eta_2.p)-\kappa'] \, .\eqno(4.19)$$
Likewise as in the previous case we obtain the formula analogical with
(4.18)
$$2\pi \kappa^{'3} i \hbar \Bigl ({\partial \over \partial x_{2}^{0}} +
\vec \sigma. {\partial \over \partial \vec x_{2}}\Bigr )\int {d^4 p\over (2\pi 
\hbar)^4} e^{{i\over \hbar}S_{21}}\, . \eqno(4.20)$$

 Finally we sketch the derived amplitudes in terms of Dirac's matrices
taken in the spinor representation.
Both the amplitudes can be written up in terms ofthe Dirac matrices
$\gamma$ in the following way
$$i\hbar \Bigl (\gamma^0{\partial \over \partial x_{2}^{0}} + \vec \gamma.
{\partial \over \partial \vec x_{2}}\Bigr ) \int {d^4 p\over (2\pi \hbar)^4}
e^{{i\over \hbar}S_{21}} $$
$$i\hbar \left(\matrix 0& {\partial \over \partial x^0}- \vec \sigma.{\partial
\over \partial \vec x_{2}} \\
 {\partial \over \partial x^{0}}+ \vec \sigma.{\partial \over \partial
 \vec x_{2}}&0 \endmatrix\right)
\int {d^4 p\over (2\pi \hbar)^4}e^{{i\over \hbar}S_{21}}$$
$$\gamma^{\mu}p_{\mu} \int {d^4 p\over (2\pi \hbar)^4}e^{{i\over \hbar}S_{21}}
\, , \eqno(4.21)$$ with the operator $p_{\mu} = i\hbar {\partial \over
\partial x^{\mu}}$. We see that both the amplitudes are defined as the
out-diagonal ones. 
    
The diagonal amplitudes are determined straightforwardly. Let us compute
the following integral
$$\int {d^4 p\over (2\pi \hbar)^4}\int {d^3 \xi_1\over \xi_{1}^{0}}\int
{d^3 \eta_{2}\over \eta_{2}^{0}}\zeta_{1}\chi_{2}^{+}e^{{i\over \hbar}S_{21}}
\kappa \delta [(\xi_1.p)-\kappa].$$ $$.{1\over 2\pi}\delta [(\xi_1.\tilde 
{\eta_{2}}] \kappa'\delta [(\eta_2.p)-\kappa'] \, . \eqno(4.22)$$
Since it holds $\delta [(\xi_1.\tilde {\eta_2}) = \delta [(\tilde {\xi_1}.
\eta_2)]$, we have then
$$\int {d^3\eta_2\over \eta_{2}^{0}}\chi_2^{+}\delta [(\tilde {\xi_1}.\eta_2)]
\delta [(\eta_2.p)-\kappa'] =
{2\pi\chi_2^{+}\over (\tilde {\xi_1}.p)} \, ,\eqno(4.23)$$
being $\eta_{2}^{0}=\kappa'(p^0-\tilde {\vec n_1}.\vec p)^{-1}$ and $\vec
n_2=\tilde {\vec n_1}$. It can be also easily found that both the equations
are equivalent to the equations 
$${\xi_1^{\mu}\over \kappa} + {\eta_2^{\mu}\over \kappa'} = 2{p^{\mu}\over
p^{2} } . \eqno(4.24)$$
However, Eq.(4.25) is the equation from which, as it was seen, Eqs.
(4.12) have been derived. Thus if we take the first equation of (4.12) in
the form 
$$\zeta_{1}^{+}(p^0 - \vec \sigma.\vec p) = \sqrt {p^{2}} \sqrt {{\kappa\over
\kappa'}}\chi_{2}^{+} $$ and we multiply this equation by $\zeta_1$, we 
obtain as a result
$$ \zeta_1\zeta_{1}^{+}(p^0 - \vec \sigma.\vec p) = \sqrt {p^{2}} \sqrt
{{\kappa\over \kappa'}}\zeta_1\chi_{2}^{+}\, .\eqno(4.25)$$
We see thus that the total amplitudes $<\phi_{f} \mid \phi_{i}>$ given
by Eqs.(4.14) and (4.23) assert that the equations
of motion linking to (2.12) plus the constraints (2.2) and (3.4) are sufficient
to construct the Feynman continual integrals (in the $p$,$\xi$,$\eta$ space)
for the relativistic particle of 1/2 spin in the spinor form. Eq.(4.18) is the
equation equivalent to Dirac's equation, defined for the two-component spinor 
with two prescribed initial constants. Such a form of the equation was 
suggested in [4] by Feynman and Gell-Mann to characterize weak interaction
decays within the V-A Lagrangian (see also [5],[6]). It is not unexpectable
that in a similar quantization scheme one is able to yield also the Pauli
equation [7].
\bigskip
\bigskip
\centerline{\bf 5. Conclusion}
\bigskip
\bigskip
   The original Feynman approach [4] meant a compromise 
between the standard operator formalism for the spin degrees of freedom and 
the scheme of continual integrals for the orbital ones. The infirmity of such 
hybrid theory was conceded by Feynman himself (see his Nobel lecture in [8]). 
One way or another, the formulation of any relativistic dynamics for Dirac
particles in terms of the continual integrals deserves to pay attention in many
regards, without excepting quantization of AAD models others than merely
the electrodynamical ones. Our model, operating with the linear interaction
and having the degree of intrinsic relationship with the electrodynamic
interaction (at least in sense of mathematical structures [10]), seems
to suggest through the Feynmanian quantization at the same time also a way
how to approach to the delicate question of spin by Feynman, using directly
the concept of linear interaction, taken in the traditionally well understand
region of quantum theory, i.e. in the domain of weak coupling. We have shown
that dynamics of the point particle with the internal degrees of freedom,
created by AAD theory, with the canonical pair of variables $\xi$ and $\eta$,
is just able to realize quantization and to yield the standard equation of
the particle with the fixed mass and 1/2 spin value.

Naturally, here one can easily anticipate difficulties, if one
has to consider states with strong coupling. In a way, the models of
quantum AAD theory can have in this case a predictive value for current
methods of the light-cone quantization [9].
                                                              
Finally we note that the discussed model provides its verifiable force
and common sense, furthermore,
in view of it means a natural generalization of the nonrelativistic
version of Pauli's equation and brings also new considerations on
the problem of particle structure [7], close to trends, for which today
the new idea of duality [11], [12] paves the way.
\bigskip
\bigskip
\centerline{\bf References}
\bigskip
\bigskip\rm
\leftline{[1] J. Weiss, Acta Phys. Polon. {\bf B39}, 977 (1988).}
\leftline{[2] R. P. Feynman, Rev. Mod. Phys. {\bf 20}, 367 (1948).}
\leftline{[3] J. Weiss, Czech. J. Phys. {\bf B39}, 1239 (1989).}
\leftline{[4] R. P. Feynman and M. Gell-Mann, Phys. Rev. {\bf 109}
193 (1958).}
\leftline{[5] E. C. G. Sudarshan and R. Marshak, Phys. Rev. {\bf 109},
1860 (1958).}
\leftline{[6] E. D. Commins and P. H. Bucksbaum, in Weak Interactions
of Leptons and Quarks,}
\leftline{\, \quad \rm Cambridge University Press, (1983).}
\leftline{[7] J. Weiss, in An AAD model of point particle and the
Pauli equation,}
\leftline{\, \quad the following paper.}
\leftline{[8] R. P. Feynman in The character of physical laws,
MIT Press,}
\leftline{\, \quad Cambridge Ma, (1965).}
\leftline{[9] in Quantum Mechanics of Fundamental Systems {\bf 2}, ed.
by C. Teitelboim}
\leftline{\, \quad and Zanelli, Plenum Press, New York, (1988).}
\leftline{[10] J. Weiss, J. Math. Phys. {\bf 27}, 1015, 1023 (1986).}
\leftline{[11] J. Polchinski, Rev. Mod. Phys. {\bf 68}, 1245 (1996).}
\leftline{[12] E. Witten, Phys. Today, May, 25 (1997).}

\end{document}